\documentclass[journal]{IEEEtran}

\usepackage{url}
\usepackage{cite}
\usepackage{latexsym}
\usepackage{amsmath}
\usepackage{amstext,amsfonts,epsf,epsfig,pifont}
\usepackage{graphics,graphicx,bm,amsmath,multicol}

\newlength{\intwidth}

\def\Xint#1{\mathchoice
{\XXint\displaystyle\textstyle{#1}}%
{\XXint\textstyle\scriptstyle{#1}}%
{\XXint\scriptstyle\scriptscriptstyle{#1}}%
{\XXint\scriptscriptstyle\scriptscriptstyle{#1}}%
\!\int}
\def\XXint#1#2#3{{\setbox0=\hbox{$#1{#2#3}{\int}$}
\vcenter{\hbox{$#2#3$}}\kern-.5\wd0}}

\def\dashint{\Xint-}

\newcommand{\bzero}{\ensuremath{\mathbf{0}}}

\newcommand{\bg}{\ensuremath{\mathbf{g}}}

\newcommand{\bj}{\ensuremath{\mathbf{j}}}
\newcommand{\bk}{\ensuremath{\mathbf{k}}}

\newcommand{\bn}{\ensuremath{\mathbf{n}}}

\newcommand{\br}{\ensuremath{\mathbf{r}}}

\newcommand{\bx}{\ensuremath{\mathbf{x}}}

\newcommand{\bz}{\ensuremath{\mathbf{z}}}

\newcommand{\bH}{\ensuremath{\mathbf{H}}}

\newcommand{\bJ}{\ensuremath{\mathbf{J}}}

\newcommand{\bR}{\ensuremath{\mathbf{R}}}

\newcommand{\bX}{\ensuremath{\mathbf{X}}}

\newcommand{\calA}{\ensuremath{\mathcal{A}}}

\newcommand{\calE}{\ensuremath{\mathcal{E}}}

\newcommand{\calH}{\ensuremath{\mathcal{H}}}

\begin{document}

\noindent \emph{The following statements are placed here in accordance with the copyright policy of the Institute of Electrical and Electronics Engineers, Inc., available online at}
\url{http://www.ieee.org/web/publications/rights/policies.html}.\\

\noindent
Lilly, J. M. (2011).  Modulated oscillations in three dimensions.\\\indent \emph{IEEE Transactions on Signal Processing}, in press.  Published\\ \indent online August 18, 2011.  \url{doi:10.1109/TSP.2011.2164914}.\\


\noindent \copyright 2011 IEEE.  Personal use of this material is permitted. Permission from IEEE must be obtained for all other uses, in any current or future media, including reprinting/republishing this material for advertising or promotional purposes, creating new collective works, for resale or redistribution to servers or lists, or reuse of any copyrighted component of this work in other works.\\

\newpage

\title{Modulated Oscillations in Three Dimensions}
\author{Jonathan~M.~Lilly,~\IEEEmembership{Member,~IEEE}
\thanks{Manuscript received March 29, 2011; revised June 20, 2011. The associate editor coordinating the review of this manuscript and approving it for publication was Prof. Antonio Napolitano.   J. M. Lilly was supported  by awards \#0751697 and \#1031002 from the Physical Oceanography program of the United States National Science Foundation.  }
\thanks{Copyright (c) 2011 IEEE. Personal use of this material is permitted. However, permission to use this material for any other purposes must be obtained from the IEEE by sending a request to pubs-permissions@ieee.org.}
\thanks{J.~M.~Lilly is with NorthWest Research Associates, PO Box 3027, Bellevue, WA 98009, USA (e-mail: lilly@nwra.com).}}

\markboth{IEEE Transactions on Signal Processing, Accepted August 2011}{Lilly: Modulated Trivariate Oscillations}

\maketitle
\begin{abstract}
The analysis of the fully three-dimensional and time-varying polarization characteristics of a modulated trivariate, or three-component, oscillation is addressed.  The use of the analytic operator enables the instantaneous three-dimensional polarization state of any square-integrable trivariate signal to be uniquely defined.  Straightforward expressions are given which permit the ellipse parameters to be recovered from data. The notions of instantaneous frequency and instantaneous bandwidth, generalized to the trivariate case, are related to variations in the ellipse properties.  Rates of change of the ellipse parameters are found to be intimately linked to the first few moments of the signal's spectrum, averaged over the three signal components.   In particular, the trivariate instantaneous bandwidth---a measure of the instantaneous departure of the signal from a single pure sinusoidal oscillation---is found to contain five contributions:  three essentially two-dimensional effects due to the motion of the ellipse within a fixed plane, and two effects due to the motion of the plane containing the ellipse.   The resulting analysis method is an informative means of describing nonstationary trivariate signals, as is illustrated with an application to a  seismic record.
\end{abstract}

\IEEEpeerreviewmaketitle

\section{Introduction}

\IEEEPARstart{M}ODULATED trivariate or three-component oscillations are important for their physical significance.  A wide variety of wavelike phenomena are aptly described as modulated trivariate oscillations, including seismic waves, internal waves in the ocean and atmosphere, and oscillations of the electric field vector in electromagnetic radiation.  Real-world waves often appear as isolated packets, as evolving nonlinear wave trains, or as sudden events whose properties change with time---all situations involving nonstationarity.  To date interest in trivariate signals has primarily been motivated by seismic applications.  In oceanography, measurements of the three-dimensional velocity field have traditionally been rare, but recent improvements in both measuring and modeling  the three-dimensional oceanic wave field, as in  \cite{dasaro96-jaot,dasaro00-jpo} and \cite{danioux08a-jpo,danioux08b-jpo} for example, make the analysis of trivariate oscillations increasingly relevant to this field as well.  Therefore a suitable analysis method for nonstationary trivariate signals would find broad applicability across a variety of disciplines.

Complex-valued three-vectors, introduced by Gibbs in 1884 \cite{gibbs}, have long been used to describe the polarization state of an oscillation in three dimensions\footnote{Gibbs referred to complex-valued three-vectors by the now-archaic term ``bivectors'', not be confused with the bivectors of geometric algebra \cite{hestenes}.}.   Previous signal analysis works have considered the three-dimensional but time-invariant polarization of trivariate signals \cite{samson77-gjras,samson80a-gjras,park87b-jgr,anderson96-itsp,donno08-itsp}, as well as the time-varying two-dimensional polarization of bivariate signals, potentially along a set of three orthogonal planes \cite{rene86-geo,diallo06-geo,lilly06-npg,schreier08b-itsp,lilly10-itsp}.  The purpose of this paper is to enable the analysis of the fully three-dimensional \emph{instantaneous} polarization state of a modulated trivariate oscillation, and to relate the variability of the polarization state to the moments of the signal's Fourier spectrum.   This is a natural but non-trivial extension of recent work on modulated bivariate oscillations by \cite{lilly10-itsp}.

One approach to the analysis of trivariate oscillations is in terms of a frequency-dependent polarization, with key contributions found in a series of works by Samson \cite{samson73-gjras,samson77-gjras,samson80a-gjras,samson80b-jgeo,samson81-geo,samson83-gjras,samson83-ang}.  Other authors, e.g. \cite{anderson96-itsp,donno08-itsp}, have similarly modeled trivariate signals as oscillatory motions with three-dimensional but time-invariant polarizations, possibly in the presence of background noise.  Estimation of the frequency-dependent polarization state based on the multitaper spectral analysis method of Thomson \cite{thomson82-ieee} was accomplished by \cite{park87b-jgr}. There the averaging necessary to estimate the spectral matrix was accomplished with an average over taper ``eigenspectra'' rather than an explict frequency-domain smoothing.

The extension to a time- and frequency-varying three-dimensional polarization was pursued by \cite{lilly95-gji,bear97-bssa,bear99b-bssa,olhede03a-prsla,olhede03b-prsla}, by employing multiple continuous \emph{wavelets}, rather than multiple global data tapers.   However, a limitation of this approach is that the polarization is a function of the multi-component wavelet \emph{transform} and not an intrinsic property of the signal; thus the definition of polarization is basis-dependent.   The time/frequency averaging implied by the use of multiple wavelets may introduce unwanted bias into the estimate, but the extent of this bias is impossible to quantify because the polarization is not independently defined.  Furthermore, reliance on the wavelet basis to define time-varying polarization sidesteps the question as to what kind of object the signal of interest is, if it is in fact not a sinusoid.

A more compelling, and ultimately more powerful, approach is to begin with a model of the signal itself.  In the univariate case, the notion of a \emph{modulated oscillation} is made precise through the use of the analytic signal \cite{gabor46-piee,ville48-cet,vakman77-spu,boashash92a-ieee,picinbono97-itsp}.  This construction permits a unique time-varying amplitude and phase pair to be associated with any square-integrable real-valued signal, see e.g. \cite{picinbono97-itsp} and references therein.  In terms of the analytic signal, intuitive and informative time-varying functions may then be found---the \emph{instantaneous frequency} \cite{gabor46-piee,ville48-cet,vakman77-spu,boashash92a-ieee,picinbono97-itsp} and \emph{instantaneous bandwidth} \cite{cohen88-spie,cohen89-ieee,cohen}---that formally provide the contributions, at each moment in time, to the first-order and second-order global moments of the signal's Fourier spectrum.  In this way time-dependent amplitude and frequency can seen as properties \emph{of the signal}.   In noisy or contaminated environments, time-frequency localization methods such as wavelet ridge analysis \cite{delprat92-itit,mallat,lilly10-itit} can then be employed to yield superior \emph{estimates} of these well-defined signal properties.

The instantaneous description of modulated oscillations using the analytic signal, and the associated instantaneous moments, has been extended to the bivariate case by several authors \cite{rene86-geo,lilly06-npg,diallo06-geo,schreier08b-itsp,lilly10-itsp}.  The use of a pair of analytic signals permits the description of a bivariate signal as an ellipse with time-varying properties,  as appears to have first been done by Ren\'{e} et al. \cite{rene86-geo} following an application of the analytic signal to univariate seismic signals by \cite{taner79-geo}.    It is a testament to the broad relevance of these ideas that there exist two distinct lines of development: one in the geophysics community \cite{taner79-geo,rene86-geo,diallo06-geo}, and another originating in the oceanographic community \cite{lilly06-npg,lilly10-itsp} based on an earlier body of work on the stationary bivariate case \cite{gonella72-dsr,mooers73-dsr,calman78a-jpo,calman78b-jpo,hayashi79-jas}.

This paper extends the analytic signal approach to investigating instantaneous signal properties to the trivariate case.   The structure of the paper is as follows.  The unique representation of a modulated trivariate oscillation in terms of a trio of analytic signals is found in Section~\ref{section:trivariate}.  This enables any real-valued trivariate signal to be uniquely described as tracing out an ellipse, the amplitude, eccentricity, and three-dimensional orientation of which all evolve in time. Simple expressions are derived which give the time-varying ellipse properties directly in terms of the trivariate analytic signal.   In Section~\ref{section:moments}, the trivariate generalizations of instantaneous frequency and bandwidth are found and are expressed in terms of rates of change of the ellipse geometry.  It is shown that  five distinct types of evolution of the ellipse geometry can give identical spectra. Application to a seismic signal is presented in Section~\ref{section:application}, followed by a concluding discussion.

All numerical code related to this paper is made freely available to the community as a package of Matlab routines\footnote{This package, called Jlab, is available at \url{http://www.jmlilly.net}.}, as discussed in Appendix~\ref{appendix:software}.

\section{Modulated Trivariate Oscillations}\label{section:trivariate}

This section develops a representation of a modulated trivariate oscillation as the trajectory of a particle orbiting a time-varying ellipse in three dimensions.  Unique specifications of the ellipse parameters are found in terms of the analytic parts of any three real-valued signals.

\subsection{Cartesian Representation}
A set of three real-valued amplitude- and frequency modulated signals may be represented as the trivariate vector
\begin{equation}
\bx(t)\equiv \left[\begin{array}{c}
        x(t)  \\
        y(t)\\
        z(t)
      \end{array}\right]\equiv
\left[\begin{array}{c}
        a_x(t) \cos\phi_x(t) \\
        a_y(t)\cos\phi_y(t)\\
        a_z(t) \cos\phi_z(t)
      \end{array}\right]\label{trivariatereal}
\end{equation}
which is herein assumed to be zero-mean and square-integrable. The representation (\ref{trivariatereal}) is non-unique in that more than one amplitude/phase pair can be associated with each real-valued signal, see e.g. \cite{picinbono97-itsp}.  However, a unique specification of the amplitudes $a_x(t)$, $a_y(t)$, and $a_z(t)$ and phases $\phi_x(t)$, $\phi_y(t)$, and $\phi_z(t)$ may be found from combining $\bx(t)$ with its Hilbert transform
\begin{equation}
\calH \bx(t)\equiv \frac{1}{\pi}\,\dashint_{-\infty}^\infty\frac{ \bx(u)}{ t-u}\,d u
\label{analyticsignalhilbert}
\end{equation}
where ``$\dashint$'' is the Cauchy principal value integral.  The six quantities appearing on the right-hand side of (\ref{trivariatereal}) are taken to be this unique set of amplitudes and phases, called the \emph{canonical set}, which is found as follows.

Pairing the real-valued signal vector with $i=\sqrt{-1}$ times its own Hilbert transform defines the \emph{analytic signal vector}
\begin{equation}
\bx_+(t)\equiv  2\calA \bx(t)\equiv \bx(t)+i\calH\bx(t)
\label{analyticsignal}
\end{equation}
where $\calA$ is called the \emph{analytic operator} \cite{boashash92a-ieee,picinbono97-itsp}. The Fourier transform of $\bx_+(t)$ is given by
\begin{equation}
\bX_+(\omega) \equiv\int_{-\infty}^\infty e^{-i\omega t} \bx_+(t)\,dt= 2U(\omega)\bX(\omega)
\end{equation}
where $\bX(\omega)$ is the Fourier transform of $\bx(t)$ and $U(\omega)$ is the Heaviside unit step function; this follows from the frequency-domain form of the analytic operator.  The amplitudes and phases of the components of the analytic signal vector
\begin{equation}
\bx_+(t)\equiv  \begin{bmatrix}x_{+}(t)\\y_{+}(t)\\z_{+}(t)
   \end{bmatrix}\equiv\begin{bmatrix}a_x(t)e^{i\phi_x(t)} \\ a_y(t)e^{i\phi_y(t)}\\a_z(t)e^{i\phi_z(t)}
   \end{bmatrix}\label{multivariateanalytic}
\end{equation}
define the canonical set of amplitudes and phases associated with $\bx(t)$, with $a_x(t)\equiv|x_{+}(t)|$ and $\phi_x(t)\equiv \arg\left\{ x_{+}(t)\right\}$ and so forth; here ``$ \arg$'' denotes the complex argument.  The real-valued signal vector is then recovered by $\bx(t)=\Re\left\{\bx_+(t)\right\}$, where ``$\Re$'' denotes the real part.

That there is a strong physical motivation in representing a univariate modulated oscillation via the canonical amplitude and phase is now well known, see e.g. \cite{vakman77-spu,boashash92a-ieee,picinbono97-itsp,cohen}.  Among the desirable features of the canonical amplitude and phase is an intimate connection between these time-varying quantities and the Fourier-domain moments of the signal, which are made use of in Section~\ref{section:moments}.   However, the analytic signal vector describes the three signal components in isolation from each other, whereas the fact that we have grouped these time series together implies that there is a reason to believe they are somehow related.

\subsection{Ellipse Representation}
Rather than consider $\bx(t)$ as a set of three disparate signals, it is more fruitful to introduce a representation which reflects possible joint structure.   A set of three sinusoidal oscillations along the coordinate axes, each having the same period but with arbitrary amplitudes and phase offsets, traces out an ellipse in three dimensions.   This suggests that a useful representation for a {\em modulated} oscillation will be in terms of an ellipse having properties that evolve with time. 

An alternate form for the analytic signal vector, the  \emph{modulated ellipse representation}, is therefore proposed as
\begin{equation}
\bx_+(t)=e^ {i\phi (t)}\bJ_3(\alpha(t))\, \bJ_1(\beta(t))\, \bJ_3(\theta(t))\,\\
\left[\begin{array}{c}
        a(t)  \\
        -ib(t)\\
        0
      \end{array}\right]\label{trivariateanalytic}
\end{equation}
where we have introduced the rotation matrices\begin{eqnarray}
\bJ_1(\theta) &\equiv&
    \begin{bmatrix}1& 0& 0\\
   0 & \cos\theta & -\sin\theta \\
   0 & \sin\theta  & \cos\theta
   \end{bmatrix}\\
   \bJ_3(\alpha) &\equiv&
    \begin{bmatrix}\cos\alpha&-\sin\alpha & 0\\
   \sin\alpha & \cos\alpha& 0\\
   0 & 0 & 1
   \end{bmatrix}
\end{eqnarray}
about the $x$ and $z$ axes respectively.  In (\ref{trivariateanalytic}), the real-valued signal $\bx(t)=\Re\left\{\bx_+(t)\right\}$ is described as the trajectory traced out in three dimensions by a particle orbiting an ellipse with time-varying amplitude, eccentricity, and orientation.    It will be shown shortly that to each value of the analytic signal $\bx_+(t)$, one may assign a unique set of the ellipse parameters.

\begin{figure}[t!]
        \noindent\begin{center}\includegraphics[height=3.5in,angle=0]{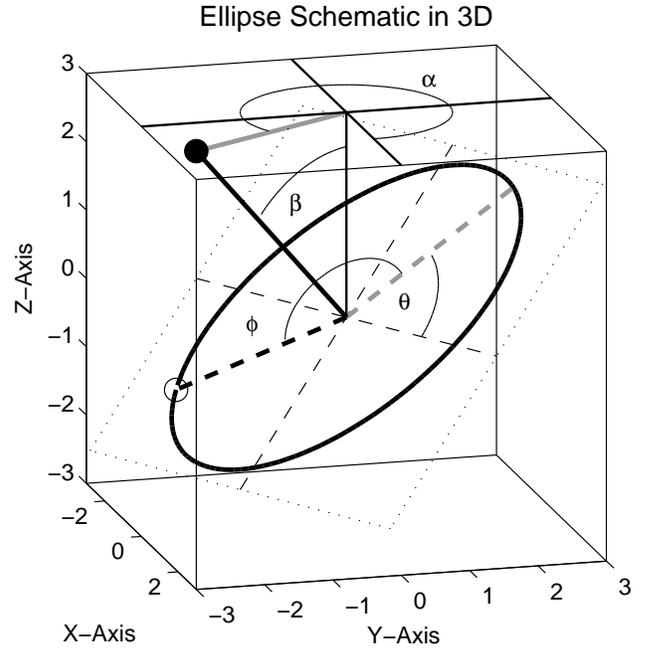}\end{center}
        \caption{\footnotesize Schematic of a modulated trivariate oscillation represented as an ellipse. A particle is shown orbiting an ellipse with constant geometry, characterized by semi-major axis $a=3$, semi-minor axis $b=2$, precession angle $\theta=\pi/3$, zenith angle $\beta=\pi/4$, and azimuth angle $\alpha=\pi/6$.  The phase increases from an initial value at $\phi=5\pi/6$, tracing out the heavy black curve through one full cycle, during which time all other ellipse parameters are constant.  The plane of the ellipse is indicated by the dotted lines, with the original $x$- and $y$-axes marked by thin dashed lines.  A heavy dashed black line, with an open circle at its end, shows position of the ``particle'' at the initial time, while a heavy gray dashed line marks the ellipse semi-major axis. A heavy solid black line, with a filled circle at its end, is the normal vector to the plane of ellipse; the projection of this vector onto the $x$--$y$ plane is shown with a heavy solid gray line at the top of the figure.
        }\label{trivariate-schematic}
\end{figure}

A sketch of an ellipse in three dimensions with all angles marked is shown in Fig.~\ref{trivariate-schematic}.  The interpretation of (\ref{trivariateanalytic}) is as follows.  An ellipse with semi-major axis $a(t)$ and semi-minor axis $b(t)$, with $a(t)\ge b(t)\ge 0$, originally lies in the $x$--$y$ plane with the major axis along the $x$-axis of the coordinate system.  The ellipse is then transformed by (i) rotating the ellipse by the \emph{precession angle} $\theta(t)$ about the $z$-axis; (ii) tilting the plane containing the ellipse about the $x$-axis by the \emph{zenith angle} $\beta(t)$; and finally (iii) rotating the normal to the plane containing the ellipse by the \emph{azimuth angle} $\alpha(t)$ about the $z$-axis. The position of a hypothetical particle along the periphery of the ellipse is specified by $\phi(t)$, called the \emph{phase angle}.  All angles are defined over  $(-\pi,\pi]$, except  for $\beta$ which is limited to $[0,\pi]$, for reasons to be discussed shortly.  Note that the rotation in three dimensions has been represented in the so-called $z$-$x$-$z$ form, as this proves convenient for the subsequent matrix multiplications.

One may replace the semi-major and semi-minor axes $a(t)$ and  $b(t)$ with two new quantities
\begin{eqnarray}
\kappa(t)&\equiv&\label{amplitudedefinition}
\sqrt{\frac{a^2(t)+b^2(t)}{2}}=\frac{1}{\sqrt{2}}\,\|\bx_+ (t)\|\\
\lambda(t) & \equiv & \frac{a^2(t)-b^2(t)}{a^2(t)+b^2(t)}\label{lambdadefinition}
\end{eqnarray}
the former being the root-mean-square ellipse amplitude, and the latter a measure of the ellipse shape; note $\|\bz\|\equiv \sqrt{\bz^H\bz}$ is the norm of some complex-valued vector $\bz$, with ``$H$'' indicating the conjugate transpose. The quantity $\lambda(t)$, like the eccentricity $\sqrt{1-b^2(t)/a^2(t)}$, varies between zero for circular motion and unity for linear motion, and thus may be termed the ellipse \emph{linearity}.\footnote{Note \cite{lilly10-itsp} defines the linearity as a signed quantity, whereas here $\lambda(t)>0$.}  Note that $\sqrt{1-\lambda^2(t)}\,\kappa^2(t)=a(t)b(t)$ gives the squared geometric mean radius of the ellipse, which could also be interpreted as a measure of the \emph{circular power}, while $\lambda(t)\kappa^2(t)=\left[a^2(t)-b^2(t)\right]/2$ could be interpreted as the \emph{linear power}.

The time variation of the signal is described by six rates of change, introduced here for future reference.  The rate of amplitude modulation is $\kappa'(t)$, while $\lambda'(t)$ is the rate of \emph{deformation} of the ellipse; here the primes indicate time derivatives. The remaining four rates of change can be said to be frequencies, in a sense, since they correspond to rates of change of angles.  The \emph{orbital frequency} $\omega_\phi(t)\equiv \phi'(t)$ gives the rate at which the particle orbits the ellipse.  The orientation of the ellipse within the plane changes at the rate $\omega_\theta(t)\equiv \theta'(t)$, which could be termed \emph{internal precession}.  This is distinguished from the azimuthal motion of the normal to the plane containing the ellipse $\omega_\alpha(t)\equiv \alpha'(t)$, or what we may call the \emph{external precession}.  To name the final quantity we may borrow a term from the description of gyroscopic motion and refer to $\omega_\beta(t)\equiv \beta'(t)$ as the rate of \emph{nutation}; as this literally means ``nodding'', it seems to appropriately describe the inward or outward motion of the normal to the plane containing the ellipse from the vertical.

\normalsize

\subsection{Comments on Angles}
In defining the angles of the ellipse representation, we constrain $0\le\beta(t)\le\pi$,  while the other three angles vary over $(-\pi,\pi]$.  These choices deserve further comment.  If the ellipse geometry is constant, i.e. only $\phi(t)$ varies in time, then the signal will repeatedly trace out the same ellipse in space, and so one should let $\phi(t)$ vary between $-\pi$ and $\pi$ to accommodate such motion.  Note that in (\ref{trivariateanalytic}) the substitutions $\theta\mapsto\theta+\pi$ and $\phi\mapsto\phi+\pi$ both have the same effect, which is to change the sign of $\bx_+(t)$.  Since $\phi(t)$ varies between $-\pi$ and $\pi$,  it might appear that $\theta$ should be limited to a range of $\pi$ in order to prevent this ambiguity; however, in practice $\theta(t)$ tends to evolve continuously in situations for which the modulated ellipse representation is suitable, and this continuity means there is no ambiguity in defining $\theta(t)$ to within a factor of $2\pi$ from moment to moment.

Clearly $\alpha(t)$, which gives the azimuth angle of the normal to the plane containing the ellipse, must vary over a range of $2\pi$ in order to allow for all orientations of the plane.  The orientation of the plane can then be completely specified with $\beta(t)$ limited between zero and $\pi/2$; however, so that the projection of the motion onto $x$--$y$ plane may be in either a clockwise or counterclockwise direction, $\beta(t)$ is allowed to vary from zero to $\pi$.  Counterclockwise motions on the $x$--$y$ plane correspond to $\beta(t)<\pi/2$, and clockwise motions to $\beta>\pi/2$.  Note that this differs from the convention of \cite{lilly10-itsp}, who in their study of bivariate modulated oscillations let $b(t)$ change sign to reflect the different directions of circulation around the ellipse.

The Hilbert transform of $\bx(t)$ decrements the phases of all Fourier components by ninety degrees, turning cosinusoids into sinusoids and sinusoids into negative cosinusoids.  Thus
\begin{equation}
\calH\bx_+(t) = -i\bx_+(t)
\end{equation}
by definition of the analytic signal. In the context of the modulated ellipse representation (\ref{trivariateanalytic}), the Hilbert transform has a simple geometric interpretation: the orbital phase $\phi(t)$ of  $\bx_+(t)$ is decremented by  $\pi/2$ with all other ellipse parameters unchanged.  Thus the signal vector and its Hilbert transform together can be used to represent a particle moving through an ellipse with fixed geometry, with $\phi(t)$ behaving as if it were a rapidly changing variable.  This is analogous to the univariate case, in which the Hilbert transform of an analytic signal $x_+(t)=a_x(t)e^{i\phi_x(t)}$ shifts the phase $\phi_x(t)$ by $\pi/2$ with the amplitude $a_x(t)$ unchanged.

\subsection{Examples}
\begin{figure*}[p!]
        \noindent\begin{center}\includegraphics[height=8.25in,angle=0]{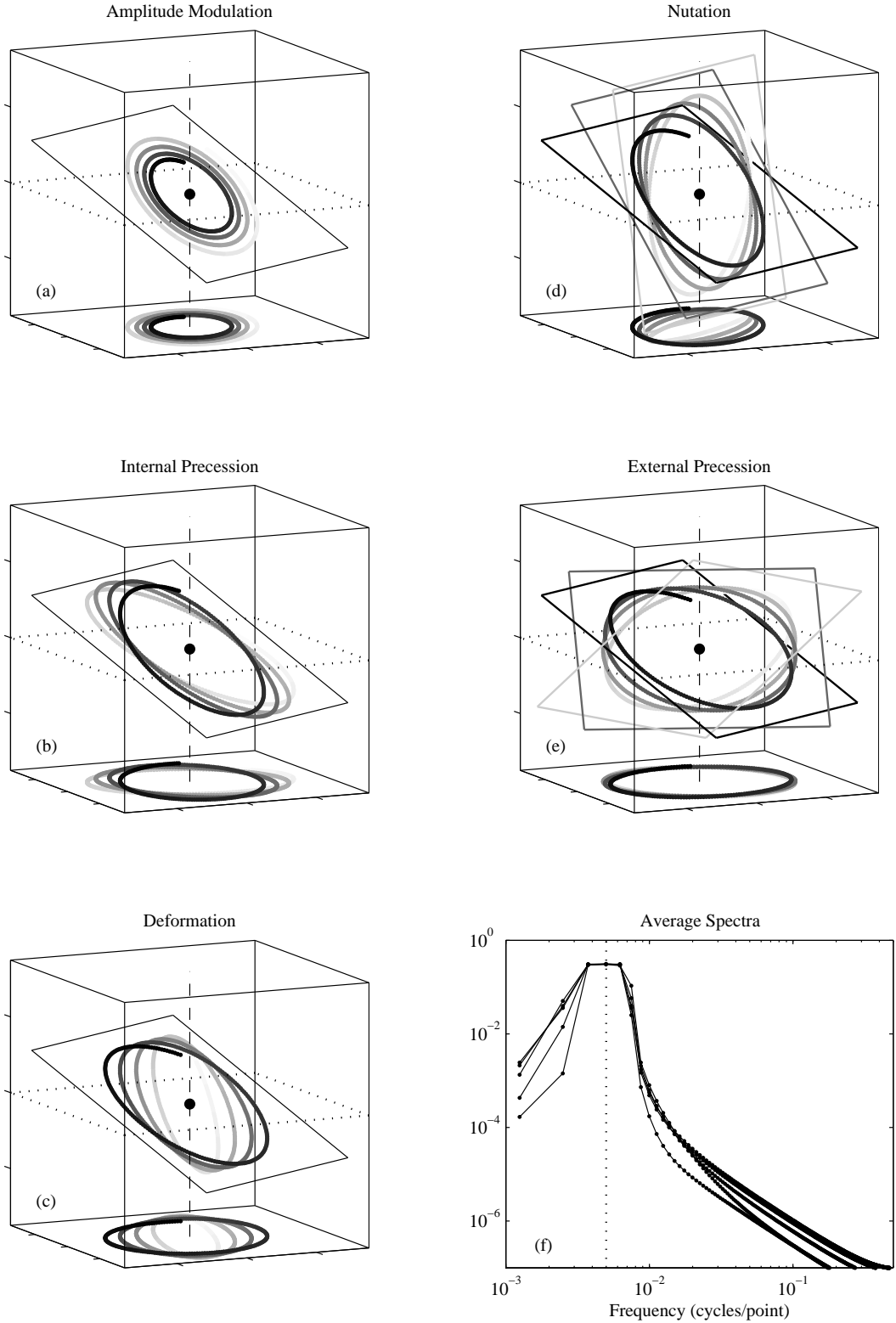}\end{center}
        \caption{\footnotesize
        Examples of modulated trivariate oscillations.  In each one of the (a--e), only one of the five parameters describing the ellipse geometry varies.  In (a--c), the plane containing the ellipse is held constant, but the ellipse amplitude $\kappa(t)$ varies in (a), the orientation $\theta(t)$ varies in (b), and the linearity $\lambda(t)$ varies in (c).  The orientation of the plane containing the ellipse varies in (d) and (e), with the zenith angle $\beta(t)$ changing in (d) and the azimuth angle $\alpha(t)$ changing in (e).  In each of these panels, a trace of the signal is shown in three dimensions with time visualized as the gray scale of the curve, with black being for early times.  The projection of the signal onto the horizontal plane is also shown.  The dashed line is a vertical line passing through the origin at $x=0$, $y=0$, while the plane $z=0$ is shown with a dotted line.  The outline of the plane (or planes) containing the ellipse is also shown in each panel. Each of these signals is 800 points in length. The remainder of the caption pertains to panel (f), which is not referred to until Section~\ref{section:moments}.  In fact the five signals in (a--e) all have been constructed to have identical mean frequencies $\overline\omega_\bx$ and mean second central moments $\overline\sigma_\bx^2$, as defined in Section~\ref{section:moments}.
        Panel (f) presents a spectral estimate of the joint spectrum $S_\bx(\omega)$ from each of the five trivariate signals, computed as described in the text. A dotted vertical line marks the global mean frequency $\overline\omega_\bx/(2\pi)=0.005$ cycles per sample point or $\pi\times 10^{-2}$ radians per sample point.  The spectra are virtually identical. }\label{trivariate-example}
\end{figure*}

To better visualize the types of signals associated with the three-dimensional ellipse representation, five examples are presented in Fig.~\ref{trivariate-example}a--e; the last panel, Fig.~\ref{trivariate-example}f, is not used until a subsequent section.  In each of the five examples, exactly one of the five rates of change describing the ellipse geometry---$\kappa'(t)$, $\lambda'(t)$, $\omega_\theta(t)$, $\omega_\alpha(t)$, and $\omega_\beta(t)$---is nonzero. As the orbital phase $\phi(t)$ varies in time along with one of the five geometry parameters, a curve is traced out in three dimensions. The projection of this motion onto the $x$--$y$ plane is also shown.  The shading of the curve represents time, with the curve being black at the initial time and fading to light gray as time progresses.

The first three examples, Fig.~\ref{trivariate-example}a--c, all involve the motion of the ellipse in a fixed plane: variation of the ellipse amplitude $\kappa(t)$ in (a), the orientation angle $\theta(t)$ in (b), and the linearity $\lambda(t)$ in (c).  These are modes of variability available to a modulated ellipse in two dimensions, as examined by \cite{lilly10-itsp}.  The last two examples reflect new possibilities due to motion of the plane containing the ellipse: tilting of the plane due to variation of $\beta(t)$ in (d), and rotation of normal vector to the plane as $\alpha(t)$ varies in (e).  This last mode can be visualized for a purely circular signal, $\lambda(t)=0$, as follows.  Imagine a plate that is spinning on a table, with a particle running around the circumference of the plate.  The spinning of the plate is associated with $\alpha(t)$, and as the plate slowly spins down $\beta(t)$ decreases to zero.

As there are six parameters in the ellipse representation (\ref{trivariateanalytic}), and also six parameters in the Cartesian representation (\ref{multivariateanalytic}) it appears reasonable to suppose that one set of parameters can be uniquely defined in terms of the other.  While it is trivial to find expressions for the Cartesian amplitudes and phases in terms of the ellipse parameters, it is not so easy to accomplish the reverse.  However, it is necessary to do so in order that the six ellipse parameters may be computed from the analytic versions of the three observed signals.  This problem is addressed in the next two subsections.

\normalsize

\subsection{The Normal Vector}

A fundamental quantity, the \emph{normal vector} to the plane containing the ellipse, will now be introduced.  The normal vector $\bn_\bx(t)$ is defined as
\begin{equation}\label{normaldef}
\bn_\bx(t)\equiv\Im\left\{\bx_+(t)\right\}\times\Re\left\{\bx_+(t)\right\}
\end{equation}
where ``$\times$'' denotes the vector product or cross product and ``$\Im$'' the imaginary part.\footnote{The symbol ``$\times$'' is also occasionally used herein to denote matrix multiplication or multiplication by a scalar, but the meaning will be clear from the context, since ``$\times$'' can only denote a cross product when a vector multiplies another vector.} That is, for two real-valued 3-vectors $\mathbf{f}=\begin{bmatrix} f_x  & f_y & f_z\end{bmatrix}^T$ and $\bg=\begin{bmatrix} g_x  & g_y & g_z\end{bmatrix}^T$, ``$T$'' being the matrix transpose, the cross product is defined as 
\begin{multline}
\mathbf{f}\times\bg\equiv 
\left(f_y g_z-f_z g_y  \right)\widehat{\mathbf{i}}\\
-\left(f_x g_z-f_z g_x  \right)\widehat\bj
+\left(f_x g_y-f_y g_x  \right)\widehat\bk
\label{crossproductdef}
\end{multline}
where $\widehat{\mathbf{i}}$, $\widehat\bj$, and $\widehat\bk$ are the unit vectors along the $x$, $y$, and $z$-axes, respectively. Note that this definition of the normal vector $\bn_\bx(t)$ to the plane containing the motion is not the same as a more familiar quantity, the angular momentum vector; the relationship between these two quantities is outside the scope of the present paper and is left to a future work.

With $\bR$ a real orthogonal matrix such that $\bR^T=\bR^{-1}$, and having unit determinant so that $\bR$ is a proper rotation matrix, the cross product transforms as
\begin{equation}\label{crossproducttransformation}
\left(\bR \mathbf{f}\right)\times \left(\bR\bg\right)=\bR\left( \mathbf{f}\times \bg\right)
\end{equation}
a result that will be used repeatedly in what follows.  This can be proven by first writing out the two sides in terms of the columns of $\bR^T$, denoted $\begin{bmatrix} \br_1 &\br_2 & \br_3\end{bmatrix}\equiv \bR^T$, which we note are related by $\epsilon_{ijk}\br_k=\br_i\times\br_j$ where $\epsilon_{ijk}$ is the Levi-Civita symbol.  Equivalence between the two sides then follows from the Binet-Cauchy identity for four real-valued $3$-vectors.

The vector $\bn_\bx(t)$ may be expressed~as, making use of  (\ref{crossproducttransformation}),
\begin{multline}
\bn_\bx(t)=\bJ_3(\alpha(t))\bJ_1(\beta(t))\bJ_3(\theta(t))\\
\begin{bmatrix}
a(t)\sin\phi(t)  \\ -b(t)\cos\phi(t)  \\ 0  \end{bmatrix}\times\begin{bmatrix}
a(t)\cos\phi(t)  \\ b(t)\sin\phi(t)  \\ 0  \end{bmatrix} \\
= a(t) b(t)\bJ_3(\alpha(t))\bJ_1(\beta(t))\widehat\bk
\end{multline}
which is oriented perpendicular to the plane containing the ellipse and has magnitude $\|\bn_\bx(t)\|=a(t) b(t)$.  Note $\pi\|\bn_\bx(t)\|$ then gives the ellipse area.  For future reference, we also define
\begin{equation}\label{unitnormal}
\widehat\bn_\bx(t)\equiv\frac{\bn_\bx(t)}{\|\bn_\bx(t)\|}=\bJ_3(\alpha(t))\bJ_1(\beta(t))\widehat\bk
\end{equation}
which is the unit normal to the plane containing the ellipse.

Since the ellipse amplitude is already known through (\ref{amplitudedefinition}), there remain five ellipse parameters to solve for.
From the normal vector one may determine three further ellipse parameters.  The linearity is found at once from
\begin{equation}
\lambda(t)= \frac{a^2(t)-b^2(t)}{a^2(t)+b^2(t)}=\sqrt{1-4\frac{\|\bn_\bx(t)\|^2}{\|\bx_+(t)\|^4}}\,.
\end{equation}
Writing out components, the normal vector becomes
\begin{equation}\label{nexpression}
\bn_\bx(t) = a(t) b(t) \begin{bmatrix}
\sin\alpha(t)\sin\beta(t) \\ -\cos\alpha(t)\sin\beta(t)  \\ \cos\beta(t) \end{bmatrix}\equiv \begin{bmatrix}
n_x(t)  \\ n_y(t)  \\ n_z(t)  \end{bmatrix}
\end{equation}
and hence the angles $\beta(t)$ and $\alpha(t)$  may be readily determined.
The former is
\begin{equation}\label{findingbeta}
\beta(t)=\Im\left\{\ln\left[n_z(t)+i \sqrt{n_x^2(t)+n_y^2(t)}\right]\right\}
\end{equation}
which recovers $0\le\beta(t)\le\pi$, while the latter is
\begin{equation}\label{findingalpha}
\alpha(t)=\Im\left\{\ln\left[-n_y(t)+in_x(t)\right]\right\}
\end{equation}
giving $-\pi<\alpha(t)\le\pi$ as desired.   The use of the ``$\Im\left\{\ln\left[\cdot\right]\right\}$'' combination amounts to the so-called four quadrant inverse tangent function, with the usual choice that $\Im\left\{\ln\left[e^{i\theta}\right]\right\}$ returns an angle $\theta$ between $-\pi$ and $\pi$. To see (\ref{findingbeta}), for example, note that\begin{multline}
\Im\left\{\ln\left[n_z(t)+i\sqrt{n_x^2(t)+n_y^2(t)}\right]\right\}=\\\Im\left\{\ln\left[\cos\beta(t)+i\left|\sin\beta(t)\right|\right]\right\}
=\Im\left\{\ln e^{i\beta(t)}\right\}=\beta(t)
\end{multline}
substituting from (\ref{nexpression}) on the second line, and using the fact that  $\left|\sin\beta(t)\right|=\sin\beta(t)$ since $0\le\beta(t)\le\pi$ by assumption.

\subsection{Precession and Phase Angles}

The values of four of the six ellipse parameters have now been established in terms of the canonical set of amplitudes and phases.  To obtain the remaining two parameters, the orientation angle $\theta(t)$ and phase angle $\phi(t)$, a representation is introduced that separates two-dimensional effects from three-dimensional effects in $\bx_+(t)$.

Let $\bH$ be the $3\times 2$ matrix which projects a 2-vector onto the $x$--$y$ plane in three dimensions, i.e.
\begin{equation}
\bH=\begin{bmatrix}1&0\\0&1\\0 &0\end{bmatrix}.
\end{equation}
Then one may write the analytic 3-vector in (\ref{trivariateanalytic}) as
\begin{equation}
\bx_+(t)
=\left[\bJ_3(\alpha(t))\bJ_1(\beta(t))\bH\right]\label{qrepresentation}
\widetilde\bx_+(t)
\end{equation}
where $\widetilde\bx_+(t)$ is a 2-vector which describes a modulated elliptical signal lying entirely within a plane.  This complex-valued 2-vector is projected into three dimensions, tilted, and then rotated to generate the analytic 3-vector $\bx_+(t)$.  Noting
\begin{multline}
\left[\bJ_3(\alpha(t))\bJ_1(\beta(t))\bH\right]^T\left[\bJ_3(\alpha(t))\bJ_1(\beta(t))\bH\right]\\=
\bH^T\bH=\begin{bmatrix}1&0\\0&1\end{bmatrix}
\label{qq}
\end{multline}
one can rearrange (\ref{qrepresentation}) to find
\begin{equation}
\widetilde\bx_+(t)
=\left[\bJ_3(\alpha(t))\bJ_1(\beta(t))\bH\right]^T\bx_+(t).
\end{equation}
As $\alpha(t)$ and $\beta(t)$ have already been determined from the previous subsection, the 2-vector $\widetilde\bx_+(t)$ is now known for any given analytic 3-vector $\bx_+(t)$.

The angles $\phi(t)$ and $\theta(t)$ may now be determined from $\widetilde\bx_+(t)$, following \cite{lilly10-itsp}. Introducing the $2\times2$ counterclockwise rotation matrix
\begin{equation}
\bJ(\theta) \equiv
    \begin{bmatrix}
   \cos\theta & -\sin\theta \\
   \sin\theta  & \cos\theta
   \end{bmatrix}
\end{equation}
the 2-vector $\widetilde\bx_+(t)$ may be expressed as
\begin{equation}\widetilde\bx_+(t)=e^{i\phi(t)}
\bJ(\theta(t))
    \begin{bmatrix}
   a(t) \\-ib(t)
   \end{bmatrix}\label{twodellipse}
\end{equation}
which is the form for a modulated elliptical signal in two dimensions examined by \cite{lilly06-npg,lilly10-itsp}.  Note that inserting (\ref{twodellipse}) into (\ref{qrepresentation}) gives (\ref{trivariateanalytic}), as required.  Now define a new 2-vector
\begin{equation}
\widetilde\bz_+(t)\equiv\begin{bmatrix}
 \widetilde a_+(t)e^{i\widetilde \phi_+(t)}\\\widetilde a_-(t)e^{i\widetilde \phi_-(t)}
\end{bmatrix}=\frac{1}{\sqrt{2}} \begin{bmatrix}
1 &i\\1 &-i
\end{bmatrix}\widetilde\bx_+(t)
\label{rotarytoCartesian}
\end{equation}
the amplitudes and phases of which are uniquely determined by the 2-vector $\widetilde\bx_+(t)$.  As discussed in \cite{lilly10-itsp}, $\widetilde\bz_+(t)$ represents the motion in two dimensions in terms of the amplitudes and phases of counterclockwise-rotating and clockwise rotating circles, and leads to simpler expressions for the ellipse parameters than does the use of $\widetilde\bx_+(t)$.

Substituting (\ref{twodellipse}) for $\widetilde\bx_+(t)$ into (\ref{rotarytoCartesian}), one finds $\widetilde\bz_+(t)$ is expressed in terms of the ellipse parameters as
 \begin{equation}
\widetilde\bz_+(t)=e^{i\phi(t)}\frac{1}{\sqrt{2}}\begin{bmatrix}
 \left[a(t)+b(t)\right]e^{i\theta(t)}\\ \left[a(t)-b(t)\right]e^{-i\theta(t)}
\end{bmatrix}
\end{equation}
and so the orientation and phase angles of the ellipse are
\begin{eqnarray}
\phi(t)&=& \left[\widetilde \phi_+(t)+\widetilde \phi_-(t)\right]/2\label{phidefinition}\\
\theta(t)&=& \left[\widetilde \phi_+(t)-\widetilde \phi_-(t)\right]/2.\label{thetadefinition}
\end{eqnarray}
All six ellipse parameters are now uniquely determined in terms of a given analytic 3-vector $\bx_+(t)$.  The functions $a(t)$, $b(t)$, $\theta(t)$, $\phi(t)$, $\alpha(t)$, and $\beta(t)$ so defined are called the \emph{canonical ellipse parameters}.

The 2-vector $\widetilde\bx_+(t)$ describes the projection of elliptical motion in three dimensions onto the plane which instantaneously contains the ellipse.  A subtle point is that $a(t)$, $b(t)$, $\theta(t)$, and $\phi(t)$ determined above are not necessarily the canonical ellipse parameters for this two-dimensional motion considered in isolation. This arises due to the fact that $\widetilde\bx_+(t)$, and similarly $\widetilde\bz_+(t)$, is not necessarily analytic. The message is that the canonical ellipse parameters give a unique description of the motion considered as a whole.  This is analogous to the key point made by \cite{picinbono97-itsp} for the univariate case that $a_x(t)e^{i\phi_x(t)}$ being analytic does not imply that $e^{i\phi_x(t)}$ is also analytic.

Note that choosing a different form for the representation of the modulated ellipse, using an alternate rotation convention such as $x$-$y$-$z$ for example, would be equivalent to (\ref{trivariateanalytic}).  The normal vector to the plane containing the ellipse, defined in (\ref{normaldef}), does not depend upon the particular ellipse representation.  Consequently in (\ref{qrepresentation}) one could have a different representation for the rotation matrix inside the square brackets, but its value must be unchanged.  The parameters
$a(t)$, $b(t)$, $\phi(t)$, and $\theta(t)$, all of which are determined by the projection of the motion onto the plane instantaneously containing the ellipse, are thus also unchanged by an alternate rotation convention.

\section{Trivariate Instantaneous Moments}\label{section:moments}

Here the first- and second-order instantaneous moments of a trivariate signal are introduced and expressed in terms of the ellipse parameters. These time-varying quantities provide the link between the ellipse parameters and the Fourier spectrum of the signal. A fundamental quantity termed the trivariate instantaneous bandwidth is seen to capture five different modes of variability of ellipse geometry, all of which contribute to the second central moment of the signal's Fourier spectrum.

\subsection{Definitions}

This section will make use of the \emph{joint instantaneous moments} of a multivariate signal introduced recently by \cite{lilly10-itsp}.  These quantities integrate to the global moments of the aggregate spectrum of a multivariate signal, just as the instantaneous moments of a univariate signal integrate to the global moments of its spectrum  \cite{gabor46-piee,ville48-cet,vakman77-spu,boashash92a-ieee,picinbono97-itsp,cohen88-spie,cohen89-ieee,cohen}.  The aggregate frequency-domain structure of the analytic vector $\bx_+(t)$ is described by the (deterministic) \emph{joint analytic spectrum}
\begin{equation}
S_\bx(\omega)\equiv \calE_\bx^ {-1}\|\bX_+(\omega)\|^ 2
\end{equation}
where the total energy of the multivariate analytic signal is
\begin{equation}
\calE_\bx\equiv \frac{1}{2\pi}\int_{0}^\infty\|\bX_+(\omega)\|^ 2d \omega=\int_{-\infty}^\infty\|\bx_+(t)\|^ 2d t.
\end{equation}
$S_\bx(\omega)$ is the average of the spectra of the $N$ analytic signals, normalized to unit energy. The \emph{joint global mean frequency}
\begin{equation}
\overline \omega_\bx  \equiv \frac{1}{2\pi}\int_{0} ^\infty \omega S_\bx(\omega)\,d\omega\label{multivariateglobalfrequency}
\end{equation}
is a measure of the average frequency content of the multivariate analytic signal $\bx_+(t)$, while the  \emph{joint global second central moment}
\begin{equation}
\overline \sigma_\bx^2  \equiv \frac{1}{2\pi}\int_{0} ^\infty (\omega-\overline \omega_\bx)^2 S_\bx(\omega)\,d\omega\label{multivariateglobalsecondmoment}
\end{equation}
gives the spread of the average spectrum about the mean frequency.  In the frequency-domain integrals above, the integration begins at zero since $\bX_+(\omega)$ has vanishing support on negative frequencies by definition.

The \emph{joint instantaneous frequency}  and \emph{joint instantaneous second central moment} are then defined by  \cite{lilly10-itsp} to be  some time-varying quantities which decompose the corresponding global moments across time, i.e. which satisfy
\begin{eqnarray}
\overline \omega_\bx  &=& \calE_\bx^ {-1}\int_{-\infty}^\infty \|\bx_+(t)\|^ 2\,\omega_\bx (t)\,d  t\label{multivariateinstantaneousmoment}\\
\overline \sigma_\bx^2  &=& \calE_\bx^ {-1}\int_{-\infty}^\infty \|\bx_+(t)\|^ 2\,\sigma_\bx^2 (t)\,d  t\label{multivariateinstantaneousmoment2}
\end{eqnarray}
noting that $\|\bx_+(t)\|^ 2$ is aggregate instantaneous power of the analytic signal vector.   Although the integrand in these expressions is non-unique, \cite{lilly10-itsp} show that the definitions
\begin{eqnarray}
\omega_\bx(t)&\equiv&\frac{\Im\left\{\bx_+^H(t)\,\bx_+'(t)\right\}}{\|\bx_+(t)\|^ 2}\label{multivariatefrequency}\\
\sigma_\bx^2(t)&\equiv&\label{multivariatesecondcentral} \frac{\left\|\bx_+'(t)-i\overline\omega_\bx\bx_+(t)\right\|^ 2}{\|\bx_+(t)\|^ 2}
\end{eqnarray}
are the natural generalizations of the standard univariate definition of the instantaneous frequency \cite{gabor46-piee} and the instantaneous second central moment \cite{cohen}. Note that (\ref{multivariatefrequency}) and (\ref{multivariatesecondcentral}) satisfy (\ref{multivariateinstantaneousmoment}) and (\ref{multivariateinstantaneousmoment2}) respectively.  Also, (\ref{multivariatesecondcentral}) is nonnegative-definite, like the global moment to which it integrates.

Thus $\omega_\bx(t)$ and $\sigma_\bx^2(t)$ can be said to give the instantaneous \emph{contributions to} the mean Fourier frequency and second central moment, respectively, or equivalently, to partition the first two Fourier moments across time.   The second-order instantaneous moment can alternately be expressed by defining the squared \emph{joint instantaneous bandwidth}
\begin{equation}\label{multivariatebandwidth}
\upsilon_\bx^2(t) \equiv  \sigma_\bx^2(t)-\left[\omega_\bx(t)-\overline\omega_\bx\right]^ 2
\end{equation}
which is that part of the instantaneous second central moment not accounted for by deviations of the instantaneous frequency from the global mean frequency.  For a univariate signal $x(t)=a_x(t)\cos\phi_x(t)$, \cite{lilly10-itsp} shows that this definition gives $\upsilon_x(t)=|a_x'(t)/a_x(t)|$, the univariate instantaneous bandwidth identified by {{\cite{cohen,cohen89-ieee,cohen88-spie}}}.    On account of the constraints  (\ref{multivariateinstantaneousmoment}) and (\ref{multivariateinstantaneousmoment2}), the scalar-valued functions $\omega_\bx(t)$ and $\upsilon_\bx(t)$ summarize the time-varying frequency content of a multivariate signal, and its spread about this frequency, in the same manner in which the standard instantaneous frequency and bandwidth would accomplish this for a univariate signal.

To find an expression for $\upsilon_\bx^2(t)$, we insert (\ref{multivariatebandwidth}) into (\ref{multivariatesecondcentral}) to give, after some manipulation,
\begin{equation}
\upsilon_\bx^2(t) =\frac{\left\|\bx_+'(t)-i \omega_\bx(t)\bx_+(t)\right\|^ 2}{\|\bx_+(t)\|^ 2}\label{twobandwidthdef}
\end{equation}
which is the normalized departure of the rate of change of the vector-valued signal from a uniform complex rotation at a single \emph{time-varying} frequency $\omega_\bx(t)$.  By contrast, $\sigma_\bx^2(t)$ is by definition the normalized departure from a uniform complex rotation at a single \emph{fixed} frequency $\overline\omega_\bx$.  The squared multivariate instantaneous bandwidth can alternately be expressed as
\begin{equation}
\upsilon_\bx^2(t)\label{alternatemultivariatebandwidth}
=\frac{\left\|\bx_+'(t)\right\|^ 2}{\|\bx_+(t)\|^ 2} -\omega_\bx^2(t)
\end{equation}
in which we have made use of the definition of the multivariate instantaneous frequency (\ref{multivariatefrequency}).  This form implies that when the modulus of the rate of change of $\bx_+(t)$
matches that expected for a set of sinusoids all locally progressing with frequency $\omega_\bx(t)$, the instantaneous bandwidth vanishes.

For the trivariate case, it is desirable to obtain expression for the instantaneous moments in terms of the ellipse parameters.  This would show how variations in the ellipse geometry contribute to the shape of the average spectrum, and at the same time provide a means for describing details of signal variation that are not resolved by the global moments.

\subsection{Trivariate Instantaneous Frequency and Bandwidth}

Forms for the trivariate instantaneous frequency and bandwidth in terms of the ellipse parameters will now be presented.  The derivations of these forms are somewhat tedious and are therefore relegated to Appendix~\ref{appendix:moments}; here we will emphasize their interpretation.  The trivariate instantaneous frequency is
\begin{equation}
\omega_\bx(t) \label{trivariatefrequency} =\overset{(\mathrm{i})}{\overbrace{\omega_\phi(t)}}+\overset{(\mathrm{ii})}{\overbrace{\sqrt{1-\lambda^2(t)}\left[\omega_\theta(t)+\omega_\alpha(t)\cos\beta(t)\right]}}
\end{equation}
and consists of two terms, (i) phase progression of the ``particle'' along the ellipse periphery and (ii) the combined effect of internal precession and external precession of the ellipse.  The squared trivariate bandwidth takes the form
\begin{multline}
\upsilon_\bx^2(t)=\label{alternatebandwidth}
\overset{(\mathrm{i})}{\overbrace{\left|\frac{\kappa'(t)}{\kappa(t)}\right|^ 2}} +
\overset{(\mathrm{ii})}{\overbrace{\frac{1}{4}\frac{\left|\lambda'(t)\right|^2}{1- \lambda^2(t)}}}\\
+\overset{(\mathrm{iii})}{\overbrace{\lambda^2(t)\left[\omega_\theta(t)+\omega_\alpha(t)\cos\beta(t)\right]^2}}
+\overset{(\mathrm{iv})}{\overbrace{\frac{\left|\widehat\bn_\bx^T(t)\,\bx_+'(t)\right|^2}{\left\| \bx_+(t)\right\|^2}}}
\end{multline}
and consists of four terms, each of which is nonnegative: (i) amplitude modulation, (ii) deformation, (iii)  precession, and (iv) the squared magnitude of that portion of the rate of change of the analytic signal that does not lie within the plane instantaneously containing the ellipse.

If the motion is contained entirely within a single plane for all time, then $\alpha(t)$ and $\beta(t)$ are constants, and $\bx_+'(t)$ has no component parallel to the normal vector $\widehat\bn_\bx(t)$.  Setting $\omega_\alpha(t)=0$ and neglecting term (iv)  in (\ref{alternatebandwidth}) recovers the bivariate forms of the instantaneous frequency and bandwidth of \cite{lilly10-itsp}
\begin{eqnarray}
\omega_\bx(t) &=& \label{bifrequency} \overset{(\mathrm{i})}{\overbrace{\omega_\phi(t)}}+\overset{(\mathrm{ii})}{\overbrace{\sqrt{1-\lambda^2(t)}\,\omega_\theta(t)}}\\
\upsilon_\bx^2(t) &=&
\overset{(\mathrm{i})}{\overbrace{\left|\frac{\kappa'(t)}{\kappa(t)}\right|^ 2}} +\overset{(\mathrm{ii})}{\overbrace{\frac{1}{4}\frac{\left|\lambda'(t)\right|^ 2}{1- \lambda^2(t)}}}+
\overset{(\mathrm{iii})}{\overbrace{\lambda^2(t)\,\omega_\theta^2(t)}}.\label{bivariatebandwidth}
\end{eqnarray}
There are  two new effects in the three-dimensional case compared with the two-dimensional case.   Firstly, in both the trivariate instantaneous frequency and bandwidth, the effect of internal precession $\omega_\theta(t)$ is modified by a term  $\omega_\alpha(t)\cos\beta(t)$ which contains the external precession rate $\omega_\alpha(t)$.   Secondly, in the trivariate instantaneous bandwidth (\ref{alternatebandwidth}), term (iv) emerges as a qualitatively new effect.

Since the external precession $\omega_\alpha(t)$ is a frequency associated with the azimuthal motion of the normal vector around the $z$-axis, it is clear that $\omega_\alpha(t)\cos\beta(t)$ is the component of external precession parallel to the normal of the plane containing the ellipse.   Thus $\omega_\theta(t)+\omega_\alpha(t)\cos\beta(t)$ could be termed the ``effective precession rate''.    When either (a) the plane containing the ellipse does not precess, so that $\omega_\alpha(t)$ vanishes, or else (b) the plane containing the ellipse rotates about a line it contains, so that $\beta(t)=\pi/2$, then the trivariate instantaneous frequency, and term (iii) in the bandwidth,  both contain no contribution from the external precession.     More generally, we may write the effective precession rate as
\begin{equation}
\omega_\theta(t)+\omega_\alpha(t)\cos\beta(t)=\label{genprecess}
\frac{\omega_\bx(t)-\omega_\phi(t)}{\sqrt{1-\lambda^2(t)}}
\end{equation}
and in this form, term (iii) of the squared instantaneous bandwidth is identical in the bivariate and trivariate cases.

\subsection{Three-Dimensional Effects in the Bandwidth}
Term (iv) in (\ref{alternatebandwidth}) involves the component of the rate of change of the analytic signal vector $\bx_+(t)$ that lies parallel to the normal vector.  That is, it is associated with the portion of changes in the real and imaginary parts of the analytic signal vector that deviate from the plane instantaneously containing the ellipse.  In Appendix~\ref{appendix:moments} this term is found to take the form
\begin{multline}\label{doesntsimplify}
\frac{\left|\widehat\bn_\bx^T(t)\bx_+'(t)\right|^2}{\left\| \bx_+(t)\right\|^2}=\\\frac{\left|\begin{bmatrix}
-\omega_\alpha(t)\sin\beta(t)& \omega_\beta(t)
\end{bmatrix}^T\widetilde\bx_+(t)\right|^2}{\|\widetilde\bx_+(t)\|^2}.
\end{multline}
Thus changes in the zenith angle of the normal vector to the ellipse with respect to the vertical, i.e. the nutation rate $\omega_\beta(t)$, as well as the component of precession of the plane containing the ellipse that lies in the horizontal plane, i.e. $\omega_\alpha(t) \sin\beta(t)$, both contribute.

Writing out (\ref{doesntsimplify}) in terms of the ellipse parameters leads to a rather complicated expression on account of the dependence of the 2-vector $\widetilde\bx_+(t)$ on the orientation angle~$\theta(t)$.  However, one can apply the Cauchy-Schwarz inequality to  (\ref{doesntsimplify}) to find
\begin{equation}\label{bound}
\frac{\left|\widehat\bn_\bx^T(t)\bx_+'(t)\right|^2}{\left\| \bx_+(t)\right\|^2}\le\omega_\alpha^2(t) \sin^2\!\beta(t)+\omega_\beta^2(t).
\end{equation}
An upper bound on the squared trivariate bandwidth is then
\begin{multline}
\upsilon_\bx^2(t)\le\label{alternatebandwidthbound}
\left|\frac{\kappa'(t)}{\kappa(t)}\right|^ 2
+\frac{1}{4}\frac{\left|\lambda'(t)\right|^ 2}{1- \lambda^2(t)}
+\omega_\beta^2(t) \\+ \left[|\omega_\theta(t)|+|\omega_\alpha(t)|\right]^2
\end{multline}
after applying the triangle inequality to term (iii) of the squared trivariate bandwidth, and setting $\lambda(t)$ to its maximum value of  unity in this term.  The bound (\ref{alternatebandwidthbound}) has a considerably simpler form than (\ref{alternatebandwidth}).  The rates of change of each of the five parameters of the ellipse geometry appear. Note that the internal and external precession rates again contribute to the same nonnegative term.

\subsection{Invariance to Coordinate Rotations}

An important point is that the terms appearing in the expressions for the trivariate instantaneous frequency $\omega_\bx(t)$ and squared instantaneous bandwidth $\upsilon_\bx^2(t)$ are independent of the choice of reference frame.  That is, replacing the original real-valued trivariate vector $\bx(t)$ with a rotated version, $\bR\bx(t)$ with $\det \bR =1$, not only preserves the values of $\omega_\bx(t)$ and $\upsilon_\bx^2(t)$, as shown by \cite{lilly10-itsp}, but also keeps the same values for the component terms.  For example,  the term $\omega_\bx(t)$ is determined from $\bx_+(t)$ and is independent of coordinate rotations, while $\omega_\bx(t)$, $\omega_\phi(t)$, and $\omega_\theta(t)$ are determined from $\widetilde\bx_+(t)$ and thus are also invariant to rotations; then (\ref{genprecess}) shows that the effective precession rate is likewise invariant.  Hence all four terms in the squared trivariate bandwidth keep the same value under coordinate rotations.  By contrast, the original definitions (\ref{multivariatefrequency}) and (\ref{multivariatebandwidth}) involve sums over terms in each signal component, and while the entire quantities $\omega_\bx(t)$ and  $\upsilon_\bx^2(t)$ are invariant to coordinate rotations, the contributing terms from each signal component are not.

\section{Applications}\label{section:application}
Two examples will be presented which illustrate the utility of the trivariate instantaneous moments.  The first example returns to the synthetic signals constructed in Fig.~\ref{trivariate-example}, while the second examines a seismic record.

\subsection{Synthetic Example}
The five signals shown in Fig.~\ref{trivariate-example}a--e have been constructed such that the joint instantaneous frequency $\omega_\bx(t)$ has identical and constant values in each panel, as does the joint instantaneous bandwidth $|\upsilon_\bx(t)|$.  Thus, the first two Fourier-domain moments of the aggregate spectra $S_\bx(\omega)$ corresponding to these five signals should be identical apart from complications arising from the finite duration of the samples.  The joint instantaneous frequency $\omega_\bx(t)$ takes a value of $\pi\times 10^{-2}$ radians per sample point, while $|\upsilon_\bx(t)|$ is set to $2.5\pi\times 10^{-4}$ radians per sample point.  The estimated aggregate spectra $S_\bx(\omega)$ associated with each of the five trivariate signals are shown in Fig.~\ref{trivariate-example}f. These are formed with a standard multitaper approach \cite{thomson82-ieee,park87a-jgr} by averaging over three ``eigenspectra'' from data tapers having a time-bandwidth product $P=2$; see \cite{park87a-jgr} for details.

It is seen that the five completely different rates of change all lead to nearly identical estimated spectra.  The estimated aggregate spectrum therefore cannot distinguish between these different types of joint structure.  However, the five rates of change can be directly computed from the observed signal, and therefore the different time-varying contributions to the Fourier bandwidth are known from the trivariate instantaneous moment analysis.  A sixth possibility also exists.  With the ellipse geometry fixed, we have $\omega_\bx(t)=\omega_\phi(t)$ and therefore $\sigma_\bx^2(t)=\left[\omega_\phi(t)-\overline\omega_\bx \right]^2$ for the first instantaneous moment and second central instantaneous moment, respectively. Thus an appropriate uniformly increasing choice of $\omega_\phi(t)$ as the particle orbits a \emph{fixed ellipse} could lead to a spectrum with the same first-order and second-order moments as in the five cases of time-varying ellipse geometry.

As a caveat to this discussion, it is worth pointing out that for the short time series segments shown in Fig.~\ref{trivariate-example}a--e, the bandwidth of the estimated spectra in Fig.~\ref{trivariate-example}f are mostly due to the taper bandwidth and not to the signal bandwidth.  Had we used much longer samples of these processes, we would have found that the five spectra differ in shape (and thus higher-order moments) despite having virtually identical first and second moments, by construction.

\subsection{Seismogram}

\begin{figure}[h!]
        \noindent\begin{center}\includegraphics[width=3.5in,angle=0]{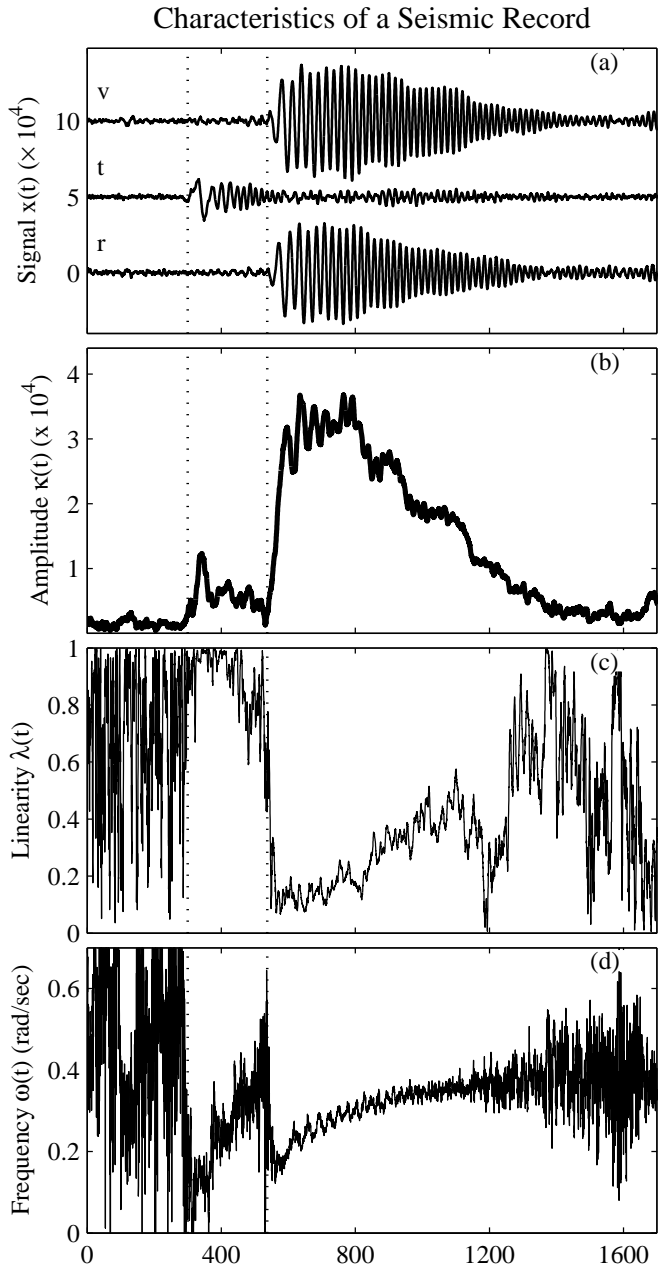}\end{center}
        \caption{\footnotesize\label{trivariate-seismic}
A three-component  seismogram from the Feb. 9, 1991, Solomon Islands earthquake as recorded in Pasadena, California.  In (a), the seismic traces are shown in the radial-transverse-vertical coordinate system. Panels (b) and (c) show the ellipse amplitude $\kappa(t)$ and linearity $\lambda(t)$, respectively, as computed from the analytic signal.  Finally (d)  presents the joint instantaneous frequency $\omega_\bx(t)$.  The $x$-axis is time in seconds from the beginning of the record. Vertical lines mark the approximate locations of the arrival times of the Love wave and the Rayleigh wave. }
\end{figure}\normalsize

As a sample dataset, the seismic trace from the Feb. 9, 1991, earthquake in the Solomon Islands as recorded at the Pasadena, California station (PAS), is presented in Fig.~\ref{trivariate-seismic}a.  This dataset is useful as an illustration since the structure is quite simple, and since it has already been examined by other authors \cite{lilly95-gji},\cite{olhede03b-prsla}.  The data is available online from the Incorporated Research Institutions for Seismology (IRIS) using the WILBER data selection interface\footnote{\url{http://www.iris.edu/dms/wilber.htm}}. The source is located at 9.93$^\circ$~S, 159.14$^\circ$~E, while the station is located at 34.15$^\circ$~N, 118.17$^\circ$~W. The bearing from the station to the source is 12.3$^\circ$ south of due west.   The $x$, $y$, and $z$ time series are rotated 12.3$^\circ$ clockwise about the vertical axis to form the radial-transverse-vertical records shown in Fig.~\ref{trivariate-seismic}a, with direction of the first (radial) axis pointing away from the seismic source.

The distinct arrivals of two different types of surface waves are clearly visible in the time series: the Love wave is a linearly polarized oscillation in the transverse direction, while the Rayleigh wave is a roughly circularly polarized wave in the radial/vertical plane. Note that the Rayleigh wave is {\em retrograde elliptical}: particle paths in this wave move towards the source when they are vertically high and away from the source when they are vertically low.  This is opposite from a gravity wave at a fluid interface, which undergoes {\em prograde elliptical} motion in the radial-vertical plane.

Taking the analytic parts of these time series, the multivariate instantaneous moments can be found at once.  The ellipse amplitude $\kappa(t)$ and linearity $\lambda(t)$, as well as the joint instantaneous frequency $\omega_\bx(t)$, are shown in  Fig.~\ref{trivariate-seismic}b--c.   The Love wave-dominated early portion of the record, between the two vertical lines, is clearly identified as being linearly polarized, while motion dominated by the Rayleigh wave after the second vertical line is associated with small linearity indicating slightly noncircular motion.   The instantaneous frequency associated with both waves is observed to increase with time.  A distinction between the frequency content of the two waves is also seen, with a sudden  drop in instantaneous frequency after the  Rayleigh wave arrival.  Near the beginning and the end of the record, rapid fluctuations of the linearity and the instantaneous frequency are consequences of a low signal-to-noise ratio.

\begin{figure*}
        \noindent\begin{center}\includegraphics[width=3.7in,angle=-90]{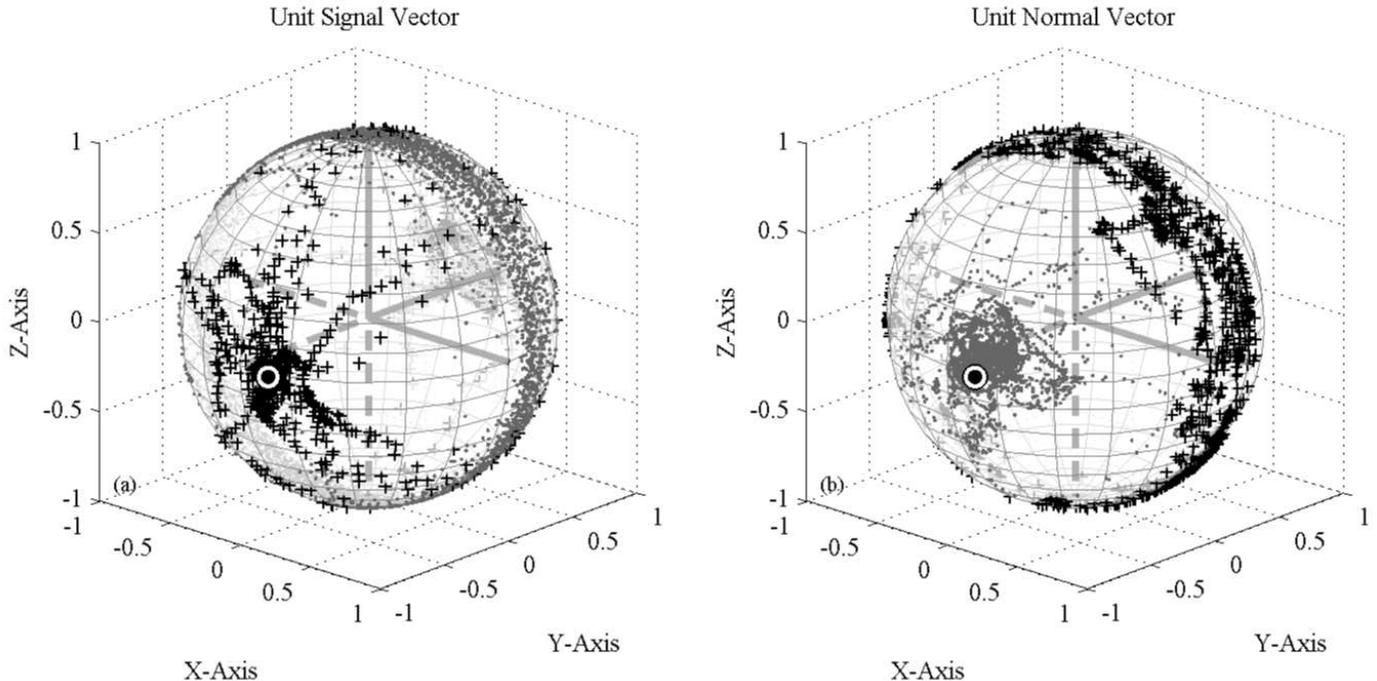}\end{center}
        \caption{\footnotesize Visualization of three-dimensional polarization of the Solomon Islands signal using (a) the real-valued unit signal vector $\widehat\bx(t)\equiv\bx(t)/\|\bx(t)\|$ and (b) the unit normal vector $\widehat\bn_\bx(t)$.  In both panels, the black ``+'' symbols show the vector position on the unit sphere during the Love wave event, i.e. between the two dotted lines in Fig.~\ref{trivariate-seismic}a, while the gray dots show the vector positions during the Rayleigh wave event, i.e. after the second dotted line. When appearing on the opposite side of the sphere, both types of symbols are plotted with a lighter shading.  The usual Cartesian coordinate system is used.  Heavy solid gray lines mark the positive $x$, $y$, and $z$ coordinate axes, while heavy dashed gray lines mark the negative axes. A black circle marks the location of the negative transverse axis at 12.3$^\circ$ east of due south.
        }\label{trivariate-sphere}
\end{figure*}

A more informative presentation of the time-varying polarization is given in Fig.~\ref{trivariate-sphere}. Here the coordinates are the standard Cartesian directions---East, North, and vertical.  In Fig.~\ref{trivariate-sphere}a, the instantaneous orientation of the real-valued signal is visualized by plotting the values of unit vector $\widehat\bx(t)\equiv\bx(t)/\|\bx(t)\|$ as points on the unit sphere. In Fig.~\ref{trivariate-sphere}b, the orientation of the unit normal vector $\widehat\bn_\bx(t)$ to the plane containing the signal and its  Hilbert transform is similarly shown.
During the Rayleigh wave, the unit normal vector offers a much more compact description of the signal. The direction of the signal vector $\widehat\bx(t)$ oscillates throughout the radial/vertical plane, whereas the unit normal vector $\widehat\bn_\bx(t)$ is quite stable
 in the positive transverse direction. Note that there is not a comparable set of points in the negative transverse direction.  This orientation of the unit normal vector indicates retrograde elliptical motion in the radial/vertical plane.  Such an orientation is expected but is difficult to visualize from the raw time series.

\normalsize

During the time period of the Love wave, the unit signal vector $\widehat\bx(t)$ oscillates between pointing in the positive transverse direction and the negative transverse direction. For such a one-dimensional signal, the plane containing the signal and its Hilbert transform is ill-defined; consequently, the direction of unit normal vector $\widehat\bn_\bx(t)$ is scattered, most likely meaninglessly, over the radial/vertical plane.  This illustrates that the ellipse parameters should be interpreted carefully for signals that are nearly one-dimensional.

\subsection{Further Possibilities}

The trivariate instantaneous moments can be applied directly to a time series to obtain useful information about the time-dependent polarization and evolution.  This works as an initial analysis step when the signal is not too structured and when the signal-to-noise ratio is sufficiently large.  These quantities can also be used as a building block in a more sophisticated analysis of realistic signals.   It is now well known that examining the instantaneous moments of a composite univariate signal, consisting of the sum of several modulated oscillations, leads to unsatisfactory results [e.g. \citen{loughlin97-sp}], and the same would be true for composite multivariate signals as well.  In general, one would like to combine the trivariate instantaneous moments with a method for extracting different modulated oscillations from an observed signal vector.   In the seismic example presented here, for example, one would like to form the trivariate instantaneous moments of the estimated Rayleigh wave signal and the estimated Love wave signal considered separately.

The instantaneous moment analysis developed herein interfaces well with the multivariate wavelet ridge analysis recently proposed by \cite{lilly09-asilomar}.  That method employs a time-frequency localization via the wavelet transform to reduce noise and to isolate different signal components from one another in a time-varying sense.  The interface between these two methods is straightforward, as one can simply take an estimated analytic signal vector from the wavelet ridge analysis and determine its ellipse parameters from the model~(\ref{trivariateanalytic}).   As real-world data is generally noisy, in most applications of the instantaneous moment analysis it will be preferable to replace the true analytic signal with one estimated from wavelet ridge analysis or some related method.

\section{Conclusions}
 
Any real-valued trivariate signal can be described as the trajectory traced out by a particle orbiting an ellipse, the amplitude, eccentricity, and three-dimensional orientation of which all evolve in time.  The rate at which the particle orbits the ellipse, together with the rates of change of the ellipse geometry, control the first two moments of the Fourier spectrum of the signal.   This perspective should be particularly valuable for describing signals which locally execute elliptical oscillations, but which may have broadband spectra on account of amplitude and frequency modulation---a class of signals which is expected to include many physical phenomena. 

While the link between instantaneous quantities derived from the analytic signal and the Fourier moments is well known for the standard univariate case,  the instantaneous  amplitude, frequency, and bandwidth take on geometrical interpretations in the trivariate case that enable a rich description of the signal's variability, permitting the distinction between qualitatively different types of motion.   Compared with the bivariate case, new terms emerge in both the instantaneous frequency and bandwidth due to motions of the plane containing the ellipse.   As a consequence, there are six different ways that the spectrum of a trivariate signal, averaged over the three signal components, may have identical mean frequency and mean bandwidth, but arising from six very different pathways of evolution of the ellipse properties.

\section*{Acknowledgment}

The facilities of the IRIS Data Management System, and specifically the IRIS Data Management Center, were used for access to waveform and metadata required in this study.  The IRIS DMS is funded through the National Science Foundation and specifically the GEO Directorate through the Instrumentation and Facilities Program of the National Science Foundation under Cooperative Agreement EAR-0552316.

\appendices

\section{A Freely Distributed Software Package}\label{appendix:software}
All software associated with this paper is distributed as a part of a Matlab toolbox called Jlab, available at \url{http://www.jmlilly.net}.   The routines used here are mostly from the Jsignal and Jellipse modules of Jlab.  The analytic part of a signal can be formed with \texttt{anatrans}.  Given an analytic signal, \texttt{instfreq} constructs the instantaneous frequency and bandwidth, as well as the joint instantaneous frequency and bandwidth of a multivariate analytic signal.  An elliptical signal in two or three dimensions is created by \texttt{ellsig} given the ellipse parameters, whereas  \texttt{ellparams} recovers the ellipse parameters from a bivariate or trivariate analytic signal.  The various ellipse geometry terms in the instantaneous bivariate or trivariate bandwidth are computed by \texttt{ellband}.  Multitaper spectral analysis is implemented by \texttt{mspec} using data tapers calculated with \texttt{sleptap}.  Ellipses are plotted in two dimensions using \texttt{ellipseplot}.   Finally, \texttt{makefigs\!\_\,trivariate} generates all figures in this paper.  All routines are well-commented and many have built-in automated tests or sample figures.

\section{Expressions for the Instantaneous Moments}\label{appendix:moments}
In this appendix, the forms of the trivariate instantaneous frequency and bandwidth in terms of the ellipse parameters are derived.  Some additional notation will facilitate the derivation.  For a complex-valued 3-vector $\bz(t)$, let
\begin{eqnarray}
\bz_\parallel(t)&\equiv&\left[ \widehat\bn_\bx^T(t)\bz(t)\right]\widehat\bn_\bx(t)\\
\bz_\perp(t)&\equiv& \bz(t)-\bz_\parallel(t)
\end{eqnarray}
be the components of $\bz(t)$ instantaneously parallel to, and perpendicular to, the unit normal vector $\widehat\bn_\bx(t)$ of the ellipse.  The real and imaginary parts of $\bz_\parallel(t)$ lie along the direction of $\widehat\bn_\bx(t)$, while the real and imaginary parts of $\bz_\perp(t)$ are in the plane perpendicular to $\widehat\bn_\bx(t)$.  Note that the parallel part of the analytic signal vector vanishes,
\begin{multline}
\bx_{\parallel}(t)=\left[ \widehat\bn_\bx^T(t)\Re\left\{\bx_+(t)\right\}\right]\widehat\bn_\bx(t)
\\+i\left[\widehat\bn_\bx^T(t)\Im\left\{\bx_+(t)\right\}\right]\widehat\bn_\bx(t)=\bzero
\end{multline}
since by definition the unit normal is perpendicular to both the real and imaginary parts of $\bx_{+}(t)$; thus $\bx_+(t)=\bx_\perp(t)$.

Using the parallel and perpendicular parts, we decompose the derivative of the analytic signal vector as
\begin{equation}
 \left[\bx_+'(t)\right]_\parallel +\label{bxdexpand}
\left[\bx_+'(t)\right]_\perp\equiv\bx_\parallel'(t) + \bx_\perp'(t)
\end{equation}
where we \emph{define} the symbols $\bx_\parallel'(t)$ and $\bx_\perp'(t)$ to mean the parallel or perpendicular part of the derivative of $\bx_+(t)$. The action of taking the parallel or perpendicular part does not commute with the derivative, thus in general $\bx_\parallel'(t)$, the parallel part of the derivative of $\bx_+(t)$, is not the same as the derivative of the parallel part of $\bx_+(t)$.  The latter quantity is
\begin{equation}\label{productrule}
\left[\bx_\parallel(t)\right]'=\bx_\parallel'(t) + \left\{\left[\widehat\bn_\bx'(t)\right]^T\bx_+(t)\right\}\widehat\bn_\bx(t)=\bzero
\end{equation}
but since $\bx_\parallel(t)$ vanishes hence $[\bx_\parallel(t)]'$ does also, we have
 \begin{equation}\label{productrule2}
\bx_\parallel'(t) =- \left\{\left[\widehat\bn_\bx'(t)\right]^T\bx_+(t)\right\}\widehat\bn_\bx(t)
\end{equation}
as an expression for $\bx_\parallel'(t)$ in terms of the rate of change of the unit normal vector $\widehat\bn_\bx(t)$.

To find expressions for the rates of change of $\bx_+(t)$ and $\widehat\bn_\bx(t)$ in terms of the ellipse parameters, note that
\begin{multline}\label{bigmatrix}
\left[\bJ_3(\alpha(t))\bJ_1(\beta(t))\right]'=
\bJ_3(\alpha(t))\bJ_1(\beta(t))\times\\\left\{
\omega_\alpha(t)\begin{bmatrix}
0 & -\cos\beta(t) & \sin\beta(t) \\
\cos\beta(t) & 0&0 \\
-\sin\beta(t) & 0 & 0
\end{bmatrix}\right.\\\left.
+\omega_\beta(t)\begin{bmatrix}
0 & 0& 0 \\
0 & 0&-1 \\
0 & 1 & 0
\end{bmatrix}\right\}
\end{multline}
as may be verified by direct computation.  From (\ref{unitnormal}), we have
\begin{equation}\label{nprime}
\widehat\bn_\bx'(t)=\bJ_3(\alpha(t))\bJ_1(\beta(t))\begin{bmatrix}
\omega_\alpha(t)\sin\beta(t) \\
-\omega_\beta(t)\\0
\end{bmatrix}
\end{equation}
for the rate of change of the unit normal vector,  which can have no component parallel to $\widehat\bn_\bx(t)$.  Using (\ref{productrule2}) then gives
\begin{equation}
\bx_\parallel'(t)\label{xprimeparallel}
=\left(\begin{bmatrix}
-\omega_\alpha(t)\sin\beta(t)&\omega_\beta(t)
\end{bmatrix}^T\widetilde\bx_+(t)\right )\widehat\bn_\bx(t)
\end{equation}
for the parallel component of the rate of change of $\bx_+(t)$.  The perpendicular component of the rate of change of  $\bx_+(t)$ is found to be
\begin{multline}
\bx_\perp'(t)\label{xprimeperp}
=\bJ_3(\alpha(t))\bJ_1(\beta(t))\bH\times\\
\left\{\widetilde\bx_+'(t)+\omega_\alpha(t)\cos\beta(t)\bJ\widetilde\bx_+(t)\right\}
\end{multline}
where we let $\bJ\equiv\bJ(\pi/2)$ with no angle argument be the $2\times 2$ ninety degree rotation matrix; (\ref{xprimeperp}) 
is obtained by differentiating  $\bx_+(t)$ as expressed by (\ref{qrepresentation}) and then using (\ref{bigmatrix}).

To simplify the expression for $\bx_\perp'(t)$, introduce for notational convenience the two-vector
\begin{equation}
\br(t)\equiv\begin{bmatrix}a(t)\\-ib(t)\end{bmatrix} = \kappa(t)\begin{bmatrix}\sqrt{1+\lambda(t)}\\-i\sqrt{1-\lambda(t)}\end{bmatrix}
\end{equation}
and then quadratic forms involving $\br(t)$ may be readily verified
\begin{eqnarray}
\br^H(t)\br(t)& =& a^2(t)+b^2(t)\label{firstqform}\\
\br^H(t) \br^*(t) & =& a^2(t)-b^2(t)\\
\br^H(t) \bJ \br(t) & =& 2 ia(t)b(t)\label{thirdqform}\\
\br^H(t) \bJ \br^*(t) & =& 0\label{lastqform}
\end{eqnarray}
which will be used shortly.  Also we find
\begin{equation}\label{rdiff}
\br'(t)=\frac{\kappa'(t)}{\kappa(t)} \br(t) + i\frac{1}{2}\frac{\lambda'(t)}{\sqrt{1-\lambda^2(t)}}\,\bJ\br^*(t)
\end{equation}
for the time derivative of $\br(t)$, using the definitions (\ref{amplitudedefinition})  and (\ref{lambdadefinition})  of the ellipse amplitude $\kappa(t)$ and linearity $\lambda(t)$.

The rate of change of the two-vector
$\widetilde\bx_+(t)$   appearing in (\ref{xprimeperp})  is given by
\begin{multline}\label{xtildederivdef}
\widetilde\bx_+'(t)=e^{i\phi(t)}\bJ(\theta(t))\times\\ \left\{\br'(t)+i\omega_\phi(t)\br(t)
+\omega_\theta(t)\bJ\br(t)
\right\}
\end{multline}
as we find by differentiating (\ref{twodellipse}).   Here we have made use of \begin{equation}
\frac{d}{dt}\,\bJ\left(\theta(t)\right)=\omega_\theta(t)\,\bJ\left(\theta(t)+\pi/2\right)= \omega_\theta(t)\,\bJ\left(\theta(t)\right)\bJ
\end{equation}
for the derivative of the $2 \times 2$ rotation matrix.  One finds
\begin{multline}\label{xtildederivdef2}
\widetilde\bx_+'(t)=e^{i\phi(t)}\bJ(\theta(t))\times \left\{\left[\frac{\kappa'(t)}{\kappa(t)} +i\omega_\phi(t)\right]\br(t)\right.\\\left.
+\omega_\theta(t)\bJ\br(t)+ i\frac{1}{2}\frac{\lambda'(t)}{\sqrt{1-\lambda^2(t)}}\,\bJ\br^*(t)
\right\}
\end{multline}
after making use of (\ref{rdiff}) for $\br'(t)$.

In terms of the parallel and perpendicular components of $\bx_+'(t)$, the instantaneous frequency and bandwidth become
\begin{eqnarray}
\omega_\bx(t)&=&\frac{\Im\left\{\bx_+^H(t)\bx_\perp'(t)\right\}}{\|\bx_+(t)\|^ 2}\label{multivariatefrequencytildeperp}\\
\upsilon_\bx^2(t)&=&\frac{\left\|\bx_\perp'(t)\right\|^ 2}{\|\bx_+(t)\|^ 2}+\frac{\left\|\bx_\parallel'(t)\right\|^ 2}{\|\bx_+(t)\|^ 2} -\omega_\bx^2(t)\label{multivariatebandwidthsecond}
\end{eqnarray}
using (\ref{multivariatebandwidth}) for the latter, and noting  $\bx_+^H(t)\bx_\parallel'(t)=0$.  For the instantaneous frequency, substituting (\ref{xprimeperp}) into (\ref{multivariatefrequencytildeperp}) gives
\begin{multline}
\bx_+^H(t)\bx_\perp'(t)\label{xxp}
\\=\widetilde\bx_+^H(t)\widetilde\bx_+'(t)+
\omega_\alpha(t)\cos\beta(t)\widetilde\bx_+^H(t)\bJ
\widetilde\bx_+(t)
\end{multline}
and using (\ref{firstqform})--(\ref{lastqform}) together with (\ref{xtildederivdef}), the trivariate instantaneous frequency expression  (\ref{trivariatefrequency}) then follows.  For the trivariate instantaneous bandwidth, note that the first term on the right-hand-side of (\ref{multivariatebandwidthsecond}) becomes, substituting (\ref{xprimeperp}) for $\bx_\perp'(t)$,
\begin{multline}\label{longbandwidthderivation}
\frac{\left\|\bx_\perp'(t)\right\|^ 2}{\|\bx_+(t)\|^2}
=\frac{\left\|\widetilde\bx_+'(t)\right\|^ 2}{\|\widetilde\bx_+(t)\|^2}
+\omega_\alpha^2(t)\cos^2\beta(t) \\+
2\omega_\alpha(t)\cos\beta(t)\frac{ \Re\left\{\widetilde\bx_+^H(t)\bJ^T\widetilde\bx_+'(t)\right\}}{\|\bx_+(t)\|^2}.
\end{multline}
We find using (\ref{firstqform})--(\ref{lastqform}) that
\begin{multline}\label{longbandwidthderivation3}
\frac{\left\|\widetilde\bx_+'(t)\right\|^ 2}{\|\widetilde\bx_+(t)\|^2}
=\left|\frac{d \ln \kappa(t)}{d t}\right|^ 2
+\frac{1}{1- \lambda^2(t)}\left|\frac{1}{2}\frac{d \lambda(t)}{d t}\right|^ 2\\+ \omega_\phi^2(t)+2\sqrt{1-\lambda^2(t)}\,\omega_\phi(t)\omega_\theta(t)+\omega_\theta^2(t)
\end{multline}
and similarly
\begin{equation}\label{xxprimeexpansion}
\frac{ \Re\left\{\widetilde\bx_+^H(t)\bJ^T\widetilde\bx_+'(t)\right\}}{\|\bx_+(t)\|^2} =
\omega_\theta(t)+\sqrt{1-\lambda^2(t)}\,\omega_\phi(t).
\end{equation}
Combining (\ref{longbandwidthderivation3}) and (\ref{xxprimeexpansion}) into (\ref{longbandwidthderivation}), then using this together with (\ref{trivariatefrequency}) for $\omega_\bx(t)$ in (\ref{multivariatebandwidthsecond}), cancelations occur, leading to the form of the trivariate bandwidth (\ref{alternatebandwidth}) given in the text.

\begin{biography}{Jonathan M. Lilly}
(M05) was born in Lansing, Michigan, in 1972. He received the B.S. degree in geology and geophysics from Yale University, New Haven, Connecticut, in 1994, and the M.S. and Ph.D. degrees in physical oceanography from the University of Washington (UW), Seattle, Washington, in 1997 and 2002, respectively. 

He was a Postdoctoral Researcher with the UW Applied Physics Laboratory and School of Oceanography, from 2002 to 2003, and with the Laboratoire d'Oc\'eanographie Dynamique et de Climatologie, Universit\'e Pierre
et Marie Curie, Paris, France, from 2003 to 2005.  From 2005 until 2010,  he was a Research Associate with Earth and Space Research in Seattle, Washington.  In 2010 he joined NorthWest Research Associates, an employee-owned scientific research corporation in Redmond, Washington,  as a Senior Research Scientist.  His research interests are oceanic vortex structures,  time/frequency analysis methods, satellite oceanography, and wave--wave interactions. 

Dr. Lilly is a member of the American Meteorological Society and of the American Geophysical Union.
\end{biography}

\label{lastpage}

\end{document}